\documentclass{ws-p9-75x6-50}

\newcommand{\eep}{\mbox{(e,e$'$p)}}
\def\bra#1{\left\langle #1\right|}
\def\ket#1{\left| #1\right\rangle}
\begin{document}

\title{Saturation Properties of Nuclear Matter and Correlated Nucleons}

\author{W. H. Dickhoff}

\address{Laboratory of Theoretical Physics, University of Gent,\\Proeftuinstraat 86, B-9000 Gent, Belgium\\Department of Physics, Washington University, St. Louis, Missouri 63130, USA\\E-mail: wimd@wuphys.wustl.edu}

\author{E. P. Roth}

\address{Department of Physics, Washington University, St. Louis, 
Missouri 63130, USA}


\maketitle

\abstracts{
A brief overview is given of the properties of spectral
functions in finite nuclei as obtained from \eep\ experiments.
Based on recent experimental data from this reaction it is 
argued that the empirical value
of the saturation density of nuclear matter is dominated by short-range
correlations.
This observation
and the observed fragmentation and depletion of the single-particle
strength in nuclei provide the motivation for attempting
a self-consistent description of the nucleon spectral
functions with full inclusion of short-range and tensor
correlations in nuclear matter.
Results for these ``second generation'' spectral functions will be discussed
with emphasis on the consequences for the saturation properties of nuclear
matter.
Arguments are presented to clarify the obscuring role of pionic long-range
correlations in this long-standing problem.
}

\section{Introduction}
Considerable progress has been made in 
probing the limits of the nuclear mean-field picture in recent years.
The primary tool in exhibiting these limits in a quantitative fashion
has been provided by the \eep\ reaction\cite{louk,pasihu}.
The qualitative features of the single-particle (sp)
strength distribution can be understood by 
realizing that a considerable mixing occurs between hole states and two-hole 
one-particle (2h1p) states. This leads to the observed fragmentation pattern
which exhibits a single peak for valence hole states near the Fermi energy,
albeit with a reduction of the strength by about 35\%\cite{louk,pasihu}.
A broadly fragmented strength distribution 
is observed for more deeply bound states which 
is related to the strong coupling to nearby 2h1p states.
For a complete understanding one requires a global depletion of mean-field
orbitals which ranges from 10\% in light nuclei to about 15\% in heavy nuclei
and nuclear matter\cite{dimu,mudi,mupodi}.
This depletion effect must be
compensated by the admixture of high-momentum
components in the ground state.
Such high-momentum nucleons have not yet been unambiguously identified 
experimentally but they do not appear near the Fermi energy\cite{mudi}.

This information about the spectral strength
distribution provides new motivation to consider the ``energy'' or
``Koltun'' sum rule\cite{miga,kolt}.
In principle, a perfect agreement of the theoretical
strength with the experimental one, at all energies and all momenta,
must yield a correspondingly good agreement
for the energy per particle, provided three-body forces are not too important.
The importance of the contribution of high-momentum nucleons, which sofar
have not been observed directly, to the energy per 
particle has been pointed out in Ref.~\citelow{mupodi}.
In addition, we will argue in Sect.~2 that the actual value of the
nuclear saturation density is dominated by the effects of short-range (and 
tensor) correlations (SRC).
Recent experimental work supports this claim.
Based on these considerations, we propose that a renewed study of the
nuclear satuation problem is in order.
Special emphasis on SRC must be utilized and we will discard all possible
contributions of long-range correlations in nuclear matter.
First results of this computationally demanding scheme 
will be discussed in some detail in Sect.~3.
The reasons for discarding the contribution of long-range correlations in
nuclear matter are discussed in Sect.~4.
Finally, some conclusions are presented in Sect.~5.

\section{General Considerations concerning Saturation of Nuclear Matter}
Absolute spectroscopic factors
associated with quasihole states have been recently obtained
for a wide range of nuclei\cite{louk,pasihu}.
The spectroscopic factors obtained in these experiments can be directly related
to the sp Green's function of the system which is given by
\begin{eqnarray}
G(\alpha , \beta ; \omega ) & = &
\sum_m\ \frac{\bra{\Psi_0^A} a_\alpha \ket{\Psi_m^{A+1}}
\bra{\Psi_m^{A+1}} a_\beta^\dagger \ket{\Psi_0^A}}
{\omega - (E_m^{A+1} - E_0^A ) + i\eta} \nonumber \\
& + & \sum_n\ \frac{\bra{\Psi_0^A} a_\beta^\dagger \ket{\Psi_n^{A-1}}
\bra{\Psi_n^{A-1}} a_\alpha \ket{\Psi_0^A}}
{\omega - (E_0^A - E_n^{A-1} ) - i\eta} .
\label{eq:prop}
\end{eqnarray}
This Lehmann-representation of the Green's function involves the exact
eigenstates and corresponding energies of the
$A$- and $A\pm 1$-particle systems.
Both the addition and removal amplitude for a particle from (to)
the ground state of the system with $A$ particles must be considered
in Eq.~(\ref{eq:prop}).
Only the removal amplitude has direct relevance for the analysis
of the \eep\ experiments.
The spectroscopic factor for the removal of a particle in the sp
orbit $\alpha$, while leaving the remaining nucleus in state $n$,
is then given by
\begin{equation}
z_\alpha^n = \left| \bra{\Psi_n^{A-1}} a_\alpha \ket{\Psi_0^A} \right|^2 ,
\label{eq:specf}
\end{equation}
which corresponds to the contribution to the numerator of the second sum
in Eq.~(\ref{eq:prop}) of state $n$ for the case $\beta = \alpha$.
Another important quantity is the spectral
function associated with sp orbit $\alpha$.
The part corresponding to the removal of particles, or hole spectral function,
is given by
\begin{equation}
S_h(\alpha , \omega) = \sum_n\
\left| \bra{\Psi_n^{A-1}} a_\alpha \ket{\Psi_0^A} \right|^2 
\delta(\omega - (E_0^A - E_n^{A-1} ) ) 
\label{eq:spefu}
\end{equation}
and characterizes the strength distribution of
the sp state $\alpha$ as a function of energy in the
$A-1$-particle system.
From this quantity one can obtain
the occupation number 
\begin{equation}
n(\alpha) = \int_{-\infty}^{\epsilon_F} d\omega\ S_h(\alpha , \omega )
= \bra{\Psi_0^A} a_\alpha^\dagger a_\alpha \ket{\Psi_0^A} .
\label{eq:occ}
\end{equation}
Experiments on ${}^{208}{\rm Pb}$ result in a spectroscopic factor of 0.65
for the removal of the last $3s_{1/2}$ proton\cite{sihu}.
Additional information about the occupation number of this orbit can be 
obtained by analyzing elastic electron scattering cross sections of 
neighboring nuclei\cite{grab}.
The occupation number for the $3s_{1/2}$ proton orbit obtained
from this analysis is about 0.75 which is 0.1 larger than the quasihole
spectroscopic factor\cite{sihu,grab}.
A recent analysis of the \eep\ reaction on ${}^{208}{\rm Pb}$
in a wide range of missing energies and for missing momenta below 270 MeV/c
yields information on the occupation numbers of all the 
more deeply-bound proton orbitals.
The data indicate that all these deeply-bound orbits are depleted by the same
amount of about 15\%\cite{louka,bat}.
The properties of these occupation numbers in ${}^{208}$Pb suggest that
the main effect of the depletion of the mean-field orbitals is due to SRC.
This statement is based on the observation that the effect of the coupling
of hole states to low-lying collective excitations only affects occupation
numbers of states in the immediate vicinity of the Fermi energy\cite{rijs} and
the knowledge that nuclear matter momentum distributions display such
global depletion effects due to short-range and tensor 
correlations\cite{fapa,vona}.
Indeed, the now observed occupation numbers in ${}^{208}$Pb\cite{louka,bat}
were anticipated some time ago based on this
information\cite{dimu}.

Based on these observations, we will now argue 
that the actual value of the empirical saturation density
of nuclear matter is dominated by SRC. 
Elastic electron scattering from ${}^{208}{\rm Pb}$\cite{froi} 
clearly pinpoints the value of the central charge density in this nucleus.
By multiplying this number by $A/Z$ one obtains the relevant central density
of heavy nuclei, corresponding to 0.16 nucleons/${\rm fm}^3$ or $k_F = 1.33~ 
{\rm fm}^{-1}$.
Since the presence of nucleons at the center of a heavy nucleus is confined
to $s$ nucleons, and their depletion is dominated by SRC,
one may conclude that the actual value of the empirical saturation density of
nuclear matter must also be closely linked to the effects
of SRC.
While this argument is particularly appropriate for the deeply bound
$1s_{1/2}$ and $2s_{1/2}$ protons, it continues to hold for the $3s_{1/2}$ 
protons which are depleted predominantly by short-range effects (up to 15\%)
and by at most 10\% due to long-range correlations\cite{sihu,grab}.

The binding energy of nuclei or nuclear matter usually
includes only mean-field contributions to the kinetic energy when 
the calculations
are based on perturbative schemes like the hole-line expansion\cite{bald}.
With the presence of high-momentum components in the ground state it becomes
relevant to consider the real kinetic and potential energy of the system
in terms of the sp strength distributions.
This result has the general form\cite{miga,kolt}
\begin{equation}
E_0^A = \bra{\Psi_0^A} \hat{H} \ket{\Psi_0^A}
= \frac{1}{2} \sum_{\alpha\beta} \bra{\alpha} T \ket{\beta} n_{\alpha\beta}
+\frac{1}{2} \sum_\alpha \int_{-\infty}^{\epsilon_F}d\omega\ \omega 
S_h(\alpha, \omega)
\label{eq:be}
\end{equation}
in the case of only two-body interactions.
In this equation $n_{\alpha\beta}$ is the one-body density matrix element
which can be directly obtained from the sp propagator.
Obvious simplifications occur in this result for the case of nuclear matter 
due to momentum conservation.
A delicate balance exists between the repulsive kinetic-energy term
and the attractive contribution of the second term in Eq.~(\ref{eq:be})
which samples the sp strength weighted by the energy
$\omega$.
When realistic spectral distributions are used to calculate
these quantities in finite nuclei unexpected results emerge\cite{mupodi}.
Such calculations for ${}^{16}{\rm O}$ indicate that the
contribution of the
quasihole states to Eq.~(\ref{eq:be}),
comprises only 37\% of the total energy
leaving 63\% for the continuum terms that represent the spectral strength
associated with the coupling to low-energy 2h1p states.
The latter contributions exhibit the presence of high-momentum
components in the nuclear ground state.
Although these high momenta 
account for only 10\% of the particles in ${}^{16}{\rm O}$, their
contribution to the energy is extremely important.
These results demonstrate the importance of treating
the dressing of nucleons in finite nuclei in determining the binding
energy per particle.
Similar considerations for nuclear matter have been available for some time
as well\cite{vonb}.

It is therefore appropriate to conclude that a careful study of 
SRC including the full fragmentation of the sp
strength is necessary for the calculation of the energy per particle
in finite nuclei and nuclear matter.
This has the additional advantage that agreement with data from the
\eep\ reaction aimed at these high-momentum components\cite{rohe} can be
used to gauge the quality of the theoretical description.
This argument can be turned inside out by noting that an exact representation
of the spectroscopic strength must lead to the correct energy
per particle according to Eq.~(\ref{eq:be}) in the case of the
dominance of two-body interactions. Clearly this perspective can only become
complete upon the successful analysis of high-momentum components in the
\eep\ reaction\cite{rohe}.
An appropriate scheme for nuclear matter that is
commensurate with these considerations, is provided
by the self-consistent calculation of nucleon spectral functions
which include, through the self-energy, the contribution of ladder diagrams
involving the propagation of these dressed particles themselves.
Some details of this scheme together with some first results\cite{libth} will
be discussed in the next section.

\section{Self-consistently Dressed Nucleons in Nuclear Matter}
The equation that represents
the calculation of the effective interaction in nuclear matter
obtained from the sum of
all ladder diagrams while propagating fully dressed particles
is given here in the partial wave representation
\begin{eqnarray}
& {} &\bra{k}\Gamma_{LL'}^{JST}(K,\Omega)\ket{k'} = 
\bra{k}V_{LL'}^{JST}(K,\Omega)\ket{k'} \nonumber \\
& + & \sum_{L''} \int_0^\infty dq\ q^2\
\bra{k}V_{LL''}^{JST}(K,\Omega)\ket{q} g_f^{II}(q;K,\Omega)
\bra{q}\Gamma_{LL''}^{JST}(K,\Omega)\ket{k'} ,
\label{eq:lad}
\end{eqnarray}
where $k,k',$ and $q$ denote relative and $K$ the total momentum.
Discrete quantum numbers correspond to total
spin, $S$, orbital angular momentum, $L,L',L''$, and
the conserved
total angular momentum and isospin, $J$ and $T$, respectively.
The energy $\Omega$ and the total momentum $K$ are conserved and act
as parameters that characterize the effective two-body interaction in the
medium.
The critical ingredient in Eq.~(\ref{eq:lad}) is the noninteracting
propagator $g_f^{II}$ which describes the propagation of the particles
in the medium from interaction to interaction.
For fully dressed particles this propagator is given by
\begin{eqnarray}
g_f^{II}(k_1,k_2;\Omega) & = &
\int_{\epsilon_F}^\infty d\omega_1\ \int_{\epsilon_F}^\infty d\omega_2\
\frac{S_p(k_1,\omega_1) S_p(k_2,\omega_2)}
{\Omega - \omega_1 -\omega_2 +i\eta} \nonumber \\
& - &
\int_{-\infty}^{\epsilon_F} d\omega_1\ \int_{-\infty}^{\epsilon_F} d\omega_2\
\frac{S_h(k_1,\omega_1) S_h(k_2,\omega_2)}
{\Omega - \omega_1 -\omega_2 -i\eta} ,
\label{eq:gtwof}
\end{eqnarray}
where individual momenta $k_1$ and $k_2$ have been used instead
of total and relative momenta as in Eq.~(\ref{eq:lad}).
The dressing of the particles is expressed by the use of particle and
hole spectral functions, $S_p$ and $S_h$, respectively.
The particle spectral function, $S_p$, is defined as a particle
addition probability density in a similar way
as the hole spectral function in Eq.~(\ref{eq:spefu}) for removal.
These spectral functions take into account that the particles propagate
with respect to the correlated ground state incorporating the
presence of high-momentum components in the ground state.
This treatment therefore includes the correlated version of the Pauli
principle leading to substantial modifications with respect to
the Pauli principle of the free Fermi gas.
This fact suggests that this correlated version may also provide a
reasonable description at higher densities.
The propagator corresponding to the Pauli principle of the free Fermi gas
is obtained from Eq.~(\ref{eq:gtwof}) by
replacing the spectral functions by strength distributions characterized
by $\delta$-functions as follows
\begin{eqnarray}
S_p(k,\omega) & = & \theta(k-k_F) \delta(\omega-\epsilon(k)) \nonumber \\
S_h(k,\omega) & = & \theta(k_F-k) \delta(\omega-\epsilon(k)) .
\label{eq:mf}
\end{eqnarray}
This leads to the so-called Galitski-Feynman propagator including hole-hole
as well as particle-particle propagation of particles characterized
by sp energies $\epsilon(k)$.
Discarding the hole-hole propagation then yields the 
Brueckner ladder diagrams.
The effective interaction obtained by solving Eq.~(\ref{eq:lad}) using
dressed propagators can be used to construct the self-energy
of the particle. With this self-energy the Dyson equation can be solved
to generate a new incarnation of the dressed propagator.
The process can then be continued by constructing anew the dressed
but noninteracting two-particle propagator according to Eq.~(\ref{eq:gtwof}).
At this stage, one can return to the ladder equation and so on until
self-consistency is achieved for the complete Green's function
which is then legitimately called a self-consistent one.

While this scheme is easy to present in equations and words, it is quite
another matter to implement it.
The recent accomplishment of implementing this self-consistency
scheme\cite{libth} builds upon earlier approximate implementations.
The first nuclear-matter spectral functions were obtained for a semirealistic
interaction by employing mean-field propagators in the ladder
equation\cite{angels,angelsa}.
Spectral functions for the Reid interaction\cite{reid}
were obtained by still employing
mean-field propagators in the ladder equation but with the introduction of
a self-consistent gap in the sp spectrum to take into account
the pairing instabilities obtained for a realistic 
interaction\cite{vonb,brian}.
The first solution of the effective interaction using dressed propagators
was obtained by employing a parametrization of the spectral 
functions\cite{chris,dick99}.

The current implementation of the self-consistent scheme for the
propagator across the summation of all ladder diagrams
includes a parametrization of the imaginary part of the nucleon self-energy.
Employing a representation in terms of two gaussians above and two below
the Fermi energy, it is possible to accurately represent 
this self-energy as generated by the contribution of relative $S$-waves
(and including the tensor coupling to the ${}^3D_1$ channel)\cite{libth}.
Self-consistency at a density corresponding to $k_F = 1.36\ {\rm fm}^{-1}$
is achieved in about ten iteration steps, each involving a considerable
amount of computational effort\cite{libth}.
A discrete version of this scheme is being implemented successfully by the
Gent group\cite{yves,gent}.
An important result pertaining to this ``second generation'' spectral functions
is related to the emergence of a common tail
at large negative energy for different momenta\cite{libth}. Such a common 
tail was previously obtained at high energy\cite{vona} in the particle domain
as a signature of SRC.
This common tail appears to play a significant role in generating some
additional binding energy at lower densities compared to conventional
Brueckner-type calculations.
At present, results for two densities corresponding to $k_F = 1.36$ and 1.45
${\rm fm}^{-1}$ have been obtained.
\begin{figure}[t]
\begin{center}
\epsfxsize=17pc 
\epsfbox{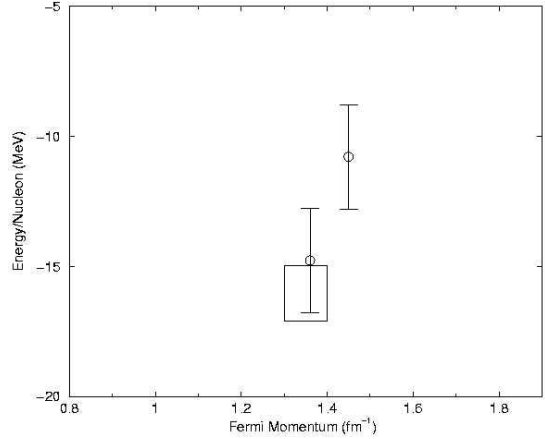} 
\caption{The energy per particle calculated at two different densities from
self-consistent spectral functions. The error bars are associated with the
lack of self-consistency related to higher order terms in $P$- and $D$-waves.
The saturation density using the Reid potential may possibly be in agreement
with the empirical result.
\label{fig:bea2}}
\end{center}
\end{figure}
Self-consistency is achieved for the contribution of the ${}^1S_0$ and 
${}^3S_1$-${}^3D_1$ channels to the self-energy. 
Higher partial waves are included in the correlated 
Hartree-Fock contribution. We have obtained additional contributions for
$L = 2$ and 3 from solutions of the dressed ladder equation after obtaining
self-consistency with the dominant $S$ waves.
The result for the
binding energy have been obtained by averaging the parametrizations of the
corresponding self-energies with and without these higher order terms for 
$L = 2$ and 3 partial waves. The difference between these two results
then provides us with a conservative estimate
of the lack of self-consistency including these terms.
This error estimate is included in Fig.~\ref{fig:bea2} for the energy per
particle calculated from the energy (Koltun) sum rule in Eq.~(\ref{eq:be}). 
These results suggest that it is possible to obtain reasonable saturation
properties for nuclear matter provided one only includes SRC.
\begin{figure}[t]
\begin{center}
\epsfxsize=20pc 
\epsfbox{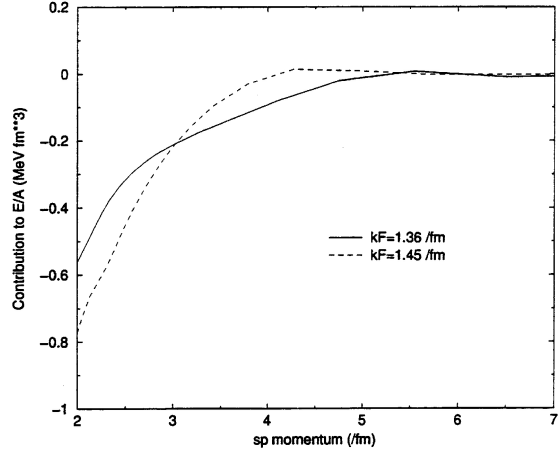} 
\caption{The high-momentum contribution to the energy per particle for
$k_F =$ 1.36 ${\rm fm}^{-1}$ (solid) and 
1.45 ${\rm fm}^{-1}$ (dashed).
This result illustrates the source of the saturation process when SRC are
considered self-consistently.
\label{fig:Image_2}}
\end{center}
\end{figure}
The special role of short-range correlations in obtaining this saturating
behavior of nuclear matter is illustrated in Fig.~\ref{fig:Image_2}.
In this figure we plot the integrand corresponding to both terms
in Eq.~(\ref{eq:be}) as a function
of momentum after performing the energy integral over the spectral function
for the same densities considered in Fig.~\ref{fig:bea2}.
It is clear from the figure that at $k_F =$ 1.36 ${\rm fm}^{-1}$ the 
high-momentum components still provide attractive contributions whereas
for $k_F =$ 1.45 ${\rm fm}^{-1}$ a changeover occurs suggesting
that at a higher density these high-momentum terms will provide
only repulsion.
From this analysis it is clear that the expected relevance of SRC in obtaining
reasonable saturation properties of nuclear matter is fully confirmed.
It remains to relate this observation to
the vast body of work on the nuclear-matter saturation
problem.

\section{Discussion of Long-Range Correlations in Nuclear Matter}
We begin this discussion by
pointing out the recent success of the Catania group in determining
the nuclear saturation curve including three hole-line 
contributions\cite{bald}.
These calculations demonstrate that a good agreement is obtained at the
three hole-line level between 
calculations that start from different prescriptions
for the auxiliary potential.
The three hole-line terms obtained in Ref.~\citelow{bald} 
indicate reasonable convergence properties compared to
the two hole-line contribution. One may therefore assume that
these results provide an accurate representation of the energy per particle
as a function of density for the case of only nonrelativistic nucleons.
The obtained saturation density corresponds to $k_F =
1.565~{\rm fm}^{-1}$ with a binding energy of -16.18 MeV.
The conclusion appears to be appropriate that additional physics
in the form of three-body forces or the inclusion of relativistic effects
is necessary to repair this obvious discrepancy with the empirical saturation
properties.

Before agreeing with this conclusion it is useful to remember that three
hole-line contributions include a third-order ring diagram characteristic
of long-range correlations.
The agreement of three hole-line calculations with advanced variational
calculations\cite{dayw} further supports the notion that important
aspects of long-range correlations are included in both these calculations.
This conclusion can also be based on the observation that hypernetted chain
calculations effectively include ring-diagram contributions
to the energy per particle although averaged over the Fermi sea\cite{jls}. 
The effect of these long-range correlations on nuclear saturation properties
is not small and can be quantified by quoting explicit results for three-
and four-body ring diagrams\cite{dfm}.
These results for the Reid potential,
including only nucleons, demonstrate that such ring-diagram terms
are dominated by attractive
contributions involving pion quantum numbers propagating
around the rings.
Furthermore, these contributions increase in importance with increasing
density.
Including the possibility of the coupling of these pionic excitation modes
to $\Delta$-hole states in 
these ring diagrams leads to an additional large increase in the binding
with increasing density\cite{dfm}.
\begin{figure}[t]
\begin{center}
\epsfxsize=11pc 
\epsfbox{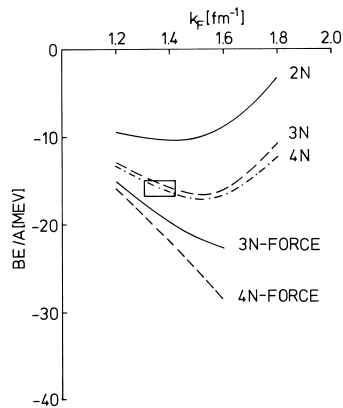} 
\caption{Contribution of three- and four-body ring diagrams involving at least
one $\Delta$-isobar configuration, to the energy per particle. 
These terms can be considered as three- and four-body force contributions
when only nucleons are considered. These contributions have been added
to the results of Ref.~\citelow{day} for the Reid interaction.
\label{fig:delt}}
\end{center}
\end{figure}
This result is illustrated in Fig.~\ref{fig:delt} where these contributions
are added to the hole-line expansion results from Ref.~\citelow{day}.
Alternatively, these terms involving $\Delta$-isobars
can also be considered as contributions due to three- and four-body
forces in the space of only nucleons.
The importance of these long-range contributions to the binding energy
is related to the possible appearance of pion condensation
at higher nuclear density.
These long-range pion-exchange dominated contributions to the
binding energy appear because of conservation of momentum in nuclear
matter. For a given momentum $q$ carried by a pion around a ring diagram,
one is able to sample coherently the attractive interaction that exists
for values of $q$ above 0.7 ${\rm fm}^{-1}$.
All ring diagrams contribute coherently when the interaction is attractive
and one may therefore obtain huge contributions at higher densities
which reflect the importance of this collective pion-propagation 
mode\cite{dickdel}.
This is clearly illustrated by the results shown in Fig.~\ref{fig:delt}.

No such collective pion-degrees of freedom are actually observed
in finite nuclei.
A substantial part of the explanation of this fact is provided by the
observation that in finite nuclei both the attractive and repulsive parts of
the pion-exchange interaction are sampled before a build-up
of long-range correlations can be achieved.
Since these contributions very nearly cancel each other, which is
further facilitated by the increased relevance of exchange terms\cite{cz1},
one does not see any marked effect on pion-like excited states in nuclei
associated with long-range pion degrees of freedom
even when $\Delta$-hole states are included\cite{cz2}.
It seems therefore reasonable to call into question the relevance
of these coherent long-range pion-exchange contributions
to the binding energy per particle since their behavior is so markedly 
different in finite and infinite systems.
Clearly, the assertion that long-range pion-exchange contributions
to the energy per particle need not be considered in explaining nuclear
saturation properties, needs to be further investigated.
In practice, this means that one must establish whether pion-exchange
in heavy nuclei already mimics the corresponding process in nuclear matter.
If this does not turn out to be the case, the arguments for considering
the nuclear-matter saturation problem only on the basis of the contribution
of SRC will be strenghtened considerably.
Furthermore, one would then also expect that the contribution of three-body
forces\cite{ppwc}
to the binding energy per particle in finite nuclei continues to be
slightly attractive when particle number is increased substantially beyond 
10\cite{wpcp}.
This point and the previous discussion also suggest that there
would be no further need for the ad-hoc repulsion added to three-body forces
used to fit nuclear-matter saturation properties\cite{cpw}. 

\section{Conclusions}
One of the critical experimental ingredients in clarifying the nature
of nuclear correlations has only become available over the last decade
and a half. It is therefore not surprising that all schemes that
have been developed to calculate nuclear-matter saturation properties
are not based on the insights that these experiments provide.
One of the aims of the present paper is to remedy this situation.
To this end we have started with
a brief reminder of experimental data obtained from the \eep\ reaction
and corresponding theoretical results,
that exhibit clear evidence that nucleons in nuclei exhibit strong correlation
effects.
Based on these considerations and the success of the theoretical calculations
to account for the qualitative features of the sp
strength distributions, it is suggested that the dressing of nucleons
must be taken into account in calculations of the energy per particle.
By identifying the dominant contribution of SRC 
to the empirical saturation density, it is argued that these correlations
need to be emphasized in the study of nuclear matter.
It is also argued that inclusion of long-range correlations, especially those
involving pion propagation, leads to an unavoidable increase in the
theoretical saturation density.
Since this collectivity in the pion channel is not observed in nuclei,
it is proposed that the corresponding correlations in nuclear matter
are not relevant for the study of nuclear saturation and should therefore
be excluded from consideration.
A scheme which fulfills this requirement and includes the propagation
of dressed particles, as required by experiment, is outlined.
Successful implementation of this scheme has recently been
demonstrated\cite{libth,gent}. First results demonstrate that these new
calculations lead to substantially lower saturation densities than 
have been obtained in the past.
The introduction of a ``nuclear-matter problem'' which focuses solely
on the contribution of SRC may therefore lead to new insight into
the long-standing problem of nuclear saturation.

\section*{Acknowledgments}
This work was supported by the U. S. National Science Foundation under Grant
No. PHY-9900713.

\end{document}